\documentclass[10pt,onecolumn,twoside]{IEEEtran}
\usepackage{calc}

\usepackage{multirow}
\usepackage{cite}
\usepackage{graphicx,subfigure}
\usepackage{psfrag}
\usepackage{amsmath,amssymb}
\usepackage{color}

\newcommand{\diag}{\mathrm{diag}}

\DeclareMathOperator*{\dotleq}{\overset{.}{\leq}}
\DeclareMathOperator*{\dotgeq}{\overset{.}{\geq}}
\DeclareMathOperator*{\defeq}{\triangleq}

\newtheorem{theorem}{Theorem}
\newtheorem{corollary}{Corollary}[theorem]

\newtheorem{lemma}{Lemma}

\newtheorem{proposition}{Proposition}

\newcommand{\bit}{\begin{itemize}}
\newcommand{\eit}{\end{itemize}}

\newcommand{\bc}{\begin{center}}
\newcommand{\ec}{\end{center}}

\newcommand{\ba}{\begin{array}}
\newcommand{\ea}{\end{array}}

\newcommand{\beq}{\begin{equation}}
\newcommand{\eeq}{\end{equation}}

\newcommand{\beqn}{\begin{equation*}}
\newcommand{\eeqn}{\end{equation*}}

\newcommand{\bean}{\begin{eqnarray*}}
\newcommand{\eean}{\end{eqnarray*}}
\newcommand{\bea}{\begin{eqnarray}}
\newcommand{\eea}{\end{eqnarray}}

\def\E{\mathbb{E}}

\def\av{\boldsymbol{a}}
\def\bv{\boldsymbol{b}}
\def\cv{\boldsymbol{c}}

\def\ev{\boldsymbol{e}}

\def\xv{\boldsymbol{x}}
\def\yv{\boldsymbol{y}}
\def\zv{\boldsymbol{z}}

\def\Hm{\boldsymbol{H}}

\def\Um{\boldsymbol{U}}
\def\Vm{\boldsymbol{V}}
\def\Wm{\boldsymbol{W}}

\newcommand{\Bc}{{\mathcal B}}

\newcommand{\T}{{\scriptscriptstyle\mathsf{T}}}
\renewcommand{\H}{{\scriptscriptstyle\mathsf{H}}}

\newcommand{\CC}{\mathbb{C}}

\newtheorem{remark}{Remark}

\renewcommand{\Bmatrix}[1]{\begin{bmatrix}#1\end{bmatrix}}

\usepackage{mathtools}
\usepackage{xifthen}

\newcommand{\cond}{\,\vert\,}

\newcommand{\deltabar}{\bar{\delta}}

\newcommand{\ya}[1][]{\ifthenelse{\isempty{#1}}{\yv_{t}^{(1)}}{\yv_{#1}^{(1)}}}
\newcommand{\yb}[1][]{\ifthenelse{\isempty{#1}}{\yv_{t}^{(2)}}{\yv_{#1}^{(2)}}}

\newcommand{\za}[1][]{\ifthenelse{\isempty{#1}}{\zv_{t}^{(1)}}{\zv_{#1}^{(1)}}}
\newcommand{\zb}[1][]{\ifthenelse{\isempty{#1}}{\zv_{t}^{(2)}}{\zv_{#1}^{(2)}}}

\newcommand{\Ha}[1][]{\ifthenelse{\isempty{#1}}{\Hm_{t}^{(1)}}{\Hm_{#1}^{(1)}}}
\newcommand{\Hb}[1][]{\ifthenelse{\isempty{#1}}{\Hm_{t}^{(2)}}{\Hm_{#1}^{(2)}}}
\newcommand{\Hi}[1][]{\ifthenelse{\isempty{#1}}{\Hm_{t}^{(i)}}{\Hm_{#1}^{(i)}}}

\newcommand{\Haa}[1][]{\ifthenelse{\isempty{#1}}{\Hm_{t}^{(11)}}{\Hm_{#1}^{(11)}}}
\newcommand{\Hab}[1][]{\ifthenelse{\isempty{#1}}{\Hm_{t}^{(12)}}{\Hm_{#1}^{(12)}}}
\newcommand{\Hba}[1][]{\ifthenelse{\isempty{#1}}{\Hm_{t}^{(21)}}{\Hm_{#1}^{(21)}}}
\newcommand{\Hbb}[1][]{\ifthenelse{\isempty{#1}}{\Hm_{t}^{(22)}}{\Hm_{#1}^{(22)}}}
\newcommand{\Hij}[1][]{\ifthenelse{\isempty{#1}}{\Hm_{t}^{(ij)}}{\Hm_{#1}^{(ij)}}}

\newcommand{\x}[1][]{\ifthenelse{\isempty{#1}}{\xv_{t}}{\xv_{#1}}}
\newcommand{\xa}[1][]{\ifthenelse{\isempty{#1}}{\xv_{t}^{(1)}}{\xv_{#1}^{(1)}}}
\newcommand{\xb}[1][]{\ifthenelse{\isempty{#1}}{\xv_{t}^{(2)}}{\xv_{#1}^{(2)}}}

\newcommand{\hHa}[1][]{\ifthenelse{\isempty{#1}}{\hat{\Hm}_{t,t'}^{(1)}}{\hat{\Hm}_{t,#1}^{(1)}}}
\newcommand{\hHb}[1][]{\ifthenelse{\isempty{#1}}{\hat{\Hm}_{t,t'}^{(2)}}{\hat{\Hm}_{t,#1}^{(2)}}}
\newcommand{\hHi}[1][]{\ifthenelse{\isempty{#1}}{\hat{\Hm}_{t,t'}^{(i)}}{\hat{\Hm}_{t,#1}^{(i)}}}

\newcommand{\hHaa}[1][]{\ifthenelse{\isempty{#1}}{\hat{\Hm}_{t,t'}^{(11)}}{\hat{\Hm}_{t,#1}^{(11)}}}
\newcommand{\hHab}[1][]{\ifthenelse{\isempty{#1}}{\hat{\Hm}_{t,t'}^{(12)}}{\hat{\Hm}_{t,#1}^{(12)}}}
\newcommand{\hHba}[1][]{\ifthenelse{\isempty{#1}}{\hat{\Hm}_{t,t'}^{(21)}}{\hat{\Hm}_{t,#1}^{(21)}}}
\newcommand{\hHbb}[1][]{\ifthenelse{\isempty{#1}}{\hat{\Hm}_{t,t'}^{(22)}}{\hat{\Hm}_{t,#1}^{(22)}}}
\newcommand{\hHij}[1][]{\ifthenelse{\isempty{#1}}{\hat{\Hm}_{t,t'}^{(ij)}}{\hat{\Hm}_{t,#1}^{(ij)}}}

\newcommand{\hHta}[1][]{\ifthenelse{\isempty{#1}}{\hat{\Hm}_{t}^{(1)}}{\hat{\Hm}_{#1}^{(1)}}}
\newcommand{\hHtb}[1][]{\ifthenelse{\isempty{#1}}{\hat{\Hm}_{t}^{(2)}}{\hat{\Hm}_{#1}^{(2)}}}

\newcommand{\hHtab}[1][]{\ifthenelse{\isempty{#1}}{\hat{\Hm}_{t}^{(12)}}{\hat{\Hm}_{#1}^{(12)}}}
\newcommand{\hHtba}[1][]{\ifthenelse{\isempty{#1}}{\hat{\Hm}_{t}^{(21)}}{\hat{\Hm}_{#1}^{(21)}}}

\newcommand{\cHta}[1][]{\ifthenelse{\isempty{#1}}{\check{\Hm}_{t}^{(1)}}{\check{\Hm}_{#1}^{(1)}}}
\newcommand{\cHtb}[1][]{\ifthenelse{\isempty{#1}}{\check{\Hm}_{t}^{(2)}}{\check{\Hm}_{#1}^{(2)}}}

\newcommand{\cHtab}[1][]{\ifthenelse{\isempty{#1}}{\check{\Hm}_{t}^{(12)}}{\check{\Hm}_{#1}^{(12)}}}
\newcommand{\cHtba}[1][]{\ifthenelse{\isempty{#1}}{\check{\Hm}_{t}^{(21)}}{\check{\Hm}_{#1}^{(21)}}}

\newcommand{\tHta}[1][]{\ifthenelse{\isempty{#1}}{\tilde{\Hm}_{t}^{(1)}}{\tilde{\Hm}_{#1}^{(1)}}}
\newcommand{\tHtb}[1][]{\ifthenelse{\isempty{#1}}{\tilde{\Hm}_{t}^{(2)}}{\tilde{\Hm}_{#1}^{(2)}}}

\newcommand{\dHta}[1][]{\ifthenelse{\isempty{#1}}{\ddot{\Hm}_{t}^{(1)}}{\ddot{\Hm}_{#1}^{(1)}}}
\newcommand{\dHtb}[1][]{\ifthenelse{\isempty{#1}}{\ddot{\Hm}_{t}^{(2)}}{\ddot{\Hm}_{#1}^{(2)}}}

\begin{document}
\sloppy

\title{Symmetric Two-User MIMO BC and IC with Evolving Feedback}
\author{Jinyuan Chen and Petros Elia
\thanks{The research leading to these results has received funding from the European Research Council under the European Community's Seventh Framework Programme (FP7/2007-2013) / EC grant agreement no. 257616 (CONECT), from the FP7 CELTIC SPECTRA project, and from Agence Nationale de la Recherche project ANR-IMAGENET.
}
\thanks{J. Chen and P. Elia are with the Mobile Communications Department, EURECOM, Sophia Antipolis, France (email: \{chenji, elia\}@eurecom.fr).}
\thanks{This paper will be presented in part at SPAWC13 \cite{CE:13spawc}.}
}

\maketitle
\thispagestyle{empty}

\begin{abstract}
Extending recent findings on the two-user MISO broadcast channel (BC) with imperfect and delayed channel state information at the transmitter (CSIT), the work here explores the performance of the two user MIMO BC and the two user MIMO interference channel (MIMO IC), in the presence of feedback with evolving quality and timeliness.
Under standard assumptions, and in the presence of $M$ antennas per transmitter and $N$ antennas per receiver, the work derives the DoF region, which is optimal for a large regime of sufficiently good (but potentially imperfect) delayed CSIT.  This region concisely captures the effect of having predicted, current and delayed-CSIT, as well as concisely captures the effect of the quality of CSIT offered at any time, about any channel.
In addition to the progress towards describing the limits of using such imperfect and delayed feedback in MIMO settings, the work offers different insights that include the fact that, an increasing number of receive antennas can allow for reduced quality feedback, as well as that no CSIT is needed for the direct links in the IC.
\end{abstract}


\section{Introduction}
\subsection{MIMO BC and MIMO IC channel models}

For the setting of the multiple-input multiple-output broadcast channel (MIMO BC), we consider the case where an $M$-antenna transmitter, sends information to two receivers with $N$ receive antennas each.  In this setting, the received signals at the two receivers take the form
\begin{subequations}\label{eq:BCmodely}
\begin{align}
\ya &= \Ha \x + \za      \label{eq:BCmodely1}\\
\yb &= \Hb \x + \zb      \label{eq:BCmodely2}
\end{align}
\end{subequations}
where $\Ha \in \CC^{N\times M} , \Hb \in \CC^{N\times M}$ respectively represent the first and second receiver channels at time $t$, where $\za,\zb$ represent unit power AWGN noise at the two receivers, where $\x \in \CC^{M\times 1 }$ is the input signal with power constraint $\E[ ||\x||^2 ] \le P$.

For the setting of the MIMO interference channel (MIMO IC), we consider a case where two transmitters, each with $M$ transmit antennas, send information to their respective receivers, each having $N$ receive antennas. In this setting, the received signals at the two receivers take the form
\begin{subequations}\label{eq:ICmodely}
\begin{align}
\ya &= \Haa \xa + \Hab \xb + \za      \label{eq:ICmodely1}\\
\yb &= \Hba \xa + \Hbb \xb + \zb    \label{eq:ICmodely2}
\end{align}
\end{subequations}
where $\Haa \in \CC^{N \times M }, \Hbb \in \CC^{N \times M }$ represent the fading matrices of the direct links of the two pairs, while $\Hab \in \CC^{N \times M }, \Hba \in \CC^{N \times M }, $ represent the fading matrices of the cross links at time $t$.

\subsection{Degrees-of-freedom as a function of feedback quality}
In the presence of perfect channel state information at the transmitter (CSIT), the degrees-of-freedom (DoF) performance\footnote{We remind the reader that in the high-SNR setting of interest, for an achievable rate pair $(R_1,R_2)$ for the first and second receiver respectively, the corresponding DoF pair $(d_1,d_2)$ is given by $d_i = \lim_{P \to \infty} \frac{R_i}{\log P},\ i=1,2$ and the corresponding DoF region is then the set of all achievable DoF pairs.} for the case of the MIMO BC, is given by (cf.~\cite{CS:03})
\begin{align}
\{d_1 &\le \min\{M,N\}, \ d_2 \le \min\{M,N\}, \ d_1+d_2 \le \min\{M,2N\}\} \label{eq:BCfull}
\end{align}
whereas for the MIMO IC, this DoF region with perfect CSIT, is given by (cf.~\cite{JF:07})
\begin{align}\label{eq:ICfull}
\{d_1 &\le \min\{M,N\} , \ d_2 \le \min\{M,N\}   , \  d_1+ d_2 \le \min\{2M, 2N, \max\{M,N\} \}\}.
\end{align}
In the absence of any CSIT though, the BC performance reduces, from that in \eqref{eq:BCfull}, to
the DoF region
\beq \{ d_1+d_2\leq \min\{M,N\} \}\eeq corresponding to a symmetric DoF corner point $(d_1=d_2=\min\{M,N\}/2)$ (cf.~\cite{HJSV:12,VV:09}).
Similarly the performance of the MIMO IC without any CSIT, reduces from the DoF region in \eqref{eq:ICfull}, to the DoF region
\beq
\{ d_1 \leq \min\{M,N\}, d_2 \leq \min\{M,N\}, d_1+d_2 \leq \min \{N,2M\} \}
\eeq corresponding to a symmetric DoF corner point $(d_1=d_2=\min\{N,2M\}/2)$ (cf. \cite{HJSV:12,VV:09}).

This gap necessitates the use of imperfect and delayed feedback, as this was studied in works like ~\cite{AGK:11o,XAJ:11b,GMK:11o,GMK:11i,VV:11t,YYGK:12,LSW:12,CE:13isit,CYE:13isit,HC:13,LSY:13,VMA:13,CE:13spawc,LHA:13,KYG:13} for specific instances.  The work here makes progress towards describing the limits of this use of imperfect and delayed feedback.

\subsection{Predicted, current and delayed CSIT}
As in \cite{CE:13it}, we consider communication of an infinite duration $n$.

For the case of the BC, we consider a random fading process $\{\Ha,\Hb\}_{t=1}^n$, and a feedback process that provides CSIT estimates $\{\hHa, \hHb\}_{t,t' = 1}^n$ (of channel $\Ha,\Hb$) at any time $t' = 1,\cdots,n$.
For the channel $\Ha,\Hb$ at a specific time $t$, the set of all available estimates $\{\hHa,\hHb\}_{t'}$, can be naturally split in the \emph{predicted estimates} $\{\hHa,\hHb\}_{t'<t}$ that are offered before the channel materializes, the \emph{current estimate} $\hHa[t],\hHb[t]$ at time $t$, and the \emph{delayed estimates} $\{\hHa,\hHb\}_{t'>t}$ that may allow for retrospective compensation for the lack of perfect quality feedback.
Naturally the fundamental measure of feedback quality is given by the precision of estimates at any time about any channel, i.e., is given by
\beq \label{eq:feedbackQuality} \{(\Ha-\hHa),(\Hb-\hHb)\}_{t,t' = 1}^n.\eeq
These estimation-error sets of course fluctuate depending on the instance of the problem, and as expected, the overall optimal performance is defined by the statistics of the above estimation errors. We here only assume that these errors have zero-mean circularly-symmetric complex Gaussian entries, that are \emph{spatially} uncorrelated, and that at any time $t$, the current estimation error is independent of the channel estimates up to that time.

Similarly for the case of the IC, we consider a fading process $\{\Haa,\Hab,\Hba,\Hbb\}_{t=1}^n$, a set of CSIT estimates $\{\hHaa,\hHab,\hHba,\hHbb\}_{t,t' = 1}^n$ and an overall feedback quality which, at any instance, is defined by
\beq \label{eq:feedbackQualityIC} \{(\Hij-\hHij)\}_{t,t' = 1}^n, \ i,j=1,2\eeq
where again, the statistics of the above error sets define the optimal performance. We will here seek to capture this relationship between performance and feedback.


\subsection{Notation, conventions and assumptions}
We will generally follow the notations and assumptions in~\cite{CE:13it}, and will adapt them to the MIMO and IC settings. When addressing the BC, we will use the notation
\beq \label{eq:alpha1}
\alpha_{t}^{(i)} =   -\lim_{P\rightarrow \infty}  \frac{  \E[||\Hi -\hHi[t]||_{F}^2]  }{  \log P } , \ \ \ \beta_{t}^{(i)}\defeq  -\lim_{P\rightarrow \infty}  \frac{ \E[||\Hi-\hHi[t+\eta] ||_{F}^2]  }{  \log P }\eeq
where $\alpha_{t}^{(i)}$ is used to describe the \emph{current quality exponent} for the CSIT for channel $\Hi$ of receiver $i, \ i=1,2$, while $\beta_{t}^{(i)}$ is used to describe the \emph{delayed quality exponents} for each user. In the above, $\eta$ can be as large as necessary, but it must be finite, as we here consider delayed CSIT that arrives after a finite delay from the channel it describes. The above used $||\bullet||_{F}$ to denote the Frobenius norm of a matrix.

Similarly when considering the MIMO IC, we will use the same notation, except that now
\beq \label{eq:alpha1IC}
\alpha_{t}^{(i)} =   -\lim_{P\rightarrow \infty}  \frac{  \E[||\Hij -\hHij[t]||_{F}^2]  }{  \log P }, \ \ \ \beta_{t}^{(i)}\defeq  -\lim_{P\rightarrow \infty}  \frac{ \E[||\Hij-\hHij[t+\eta] ||_{F}^2]  }{  \log P }, \ \ \ i\neq j \eeq
where $\alpha_{t}^{(1)},\beta_{t}^{(1)}$ will correspond to the CSIT quality for the cross link $\Hab$ where this CSIT is available at transmitter~2, and where $\alpha_{t}^{(2)},\beta_{t}^{(2)}$ will correspond to the CSIT quality for the cross link $\Hba$ where this CSIT is available at transmitter~1\footnote{When treating the IC case, emphasis is placed on the CSIT of the cross links because, as it will turn out, the DoF region will be achieved without any knowledge of the direct links.  This is a small improvement over~\cite{YYGK:12} where both transmitters were assumed to have static-quality CSIT for all the channels $(\Hba,\Hba,\Haa,\Hbb)$.}

As argued in \cite{CE:13it}, the results in \cite{Jindal:06m,Caire+:10m} easily show that without loss of generality, in the DoF setting of interest, we can restrict our attention to the range
\beq \label{eq:exponentRange} 0\leq \alpha_t^{(i)} \leq \beta_t^{(i)} \leq 1.\eeq
Here having $\alpha_t^{(1)} = \alpha_t^{(2)} = 1$, corresponds to the highest quality CSIT with perfect timing (full CSIT) for the specific channel at time $t$, while having $\beta_t^{(i)} = 1$ corresponds to having perfect delayed CSIT for the same channel, i.e., it corresponds to the case where at some point $t'>t$, the transmitter has perfect estimates of the channel that materialized at time $t$.

Furthermore we will use the notation
\beq \label{eq:averages}
\bar{\alpha}^{(i)} \defeq  \lim_{n \rightarrow \infty}\frac{1}{n}\sum^{n}_{t=1}\alpha_t^{(i)}, \quad \bar{\beta}^{(i)} \defeq \lim_{n \rightarrow \infty}\frac{1}{n}\sum^{n}_{t=1}\beta_t^{(i)}, \quad i=1,2\eeq
to denote the average of the quality exponents. As in \cite{CE:13it} we will adopt the mild assumption that any sufficiently long subsequence $\{\alpha^{(1)}_t\}_{t=\tau}^{\tau+T}$ (resp. $\{\alpha^{(2)}_t\}_{t=\tau}^{\tau+T},\{\beta^{(1)}_t\}_{t=\tau}^{\tau+T},\{\beta^{(2)}_t\}_{t=\tau}^{\tau+T}$) has an average that converges to the long term average $\bar{\alpha}^{(1)}$ (resp. $\bar{\alpha}^{(2)},\bar{\beta}^{(1)},\bar{\beta}^{(2)}$), for any $\tau$ and for some finite $T$ that can be chosen to be sufficiently large to allow for the above convergence.

Implicit in our definition of the quality exponents, is our assumption that $\E[||\Ha-\hHa[t']||_{F}^2]  \leq  \E[||\Ha-\hHa[t'']||_{F}^2], \ \E[||\Hb-\hHb[t']||_{F}^2]  \leq  \E[||\Hb-\hHb[t'']||_{F}^2] $, for any $t'>t''$, (similarly for the IC case) which simply reflects the fact that one can revert back to past estimates of statistically better quality. This assumption can be removed - after a small change in the definition of the quality exponents - without an effect to the main result.

Throughout this paper, $(\bullet)^\T$ and $(\bullet)^{\H}$ will denote the transpose and conjugate transpose of a matrix respectively, while $\diag(\bullet)$ will denote a diagonal matrix, $||\bullet||$ will denote the Euclidean norm, and $|\bullet|$ will denote the magnitude of a scalar. $o(\bullet)$ comes from the standard Landau notation, where $f(x) = o(g(x))$ implies $\lim_{x\to \infty} f(x)/g(x)=0$.  We also use $\doteq$ to denote \emph{exponential equality}, i.e., we write $f(P)\doteq P^{B}$ to denote $\displaystyle\lim_{P\to\infty}\frac{\log f(P)}{\log P}=B$.  Similarly $\dotgeq$ and $\dotleq$ will denote exponential inequalities.  Logarithms are of base~$2$.
$(\bullet)^{+}=\max\{\bullet, 0\}$.

Furthermore we adhere to the common convention (see \cite{MAT:11c,MJS:12,GJ:12o,YKGY:12d,YYGK:12}) of assuming perfect and global knowledge of channel state information at the receivers (perfect global CSIR), where the receivers know all channel states and all estimates.  We will also adopt the common convention (see \cite{KYGY:12o, YKGY:12d, GJ:12o,LSW:05}) of assuming that the current estimation error is statistically independent of current and past estimates.  A discussion on this can be found in~\cite{CE:13it} which argues that this assumption fits well with many channel models, spanning from the fast fading channel (i.i.d. in time), to the correlated channel model as this is considered in \cite{KYGY:12o}, to the quasi-static block fading model where the CSIT estimates are successively refined while the channel remains static. Additionally we consider the entries of each estimation error matrix $\Hi -\hHi$ to be i.i.d. Gaussian~\footnote{We here make it clear that we are simply referring to the $MN$ entries in each such specific matrix $\Hi -\hHi$, and that we certainly do not suggest that the error entries are i.i.d. in time or across users.}.
Finally we will refer to a CSIT process with `sufficiently good delayed CSIT', to be a process for which $\min \{ \bar{\beta}^{(1)}, \bar{\beta}^{(2)}\} \geq \min \{1, M-\min\{M,N\},\frac{ N(1 + \bar{\alpha}^{(1)} +  \bar{\alpha}^{(2)}) }{\min\{M,2N\}+N}, \frac{N( 1 +\min\{\bar{\alpha}^{(1)} +\bar{\alpha}^{(2)}  \}) }{\min\{M,2N\}}  \}$.

\subsection{Existing results directly relating to the current work}
The work here builds on the ideas of \cite{MAT:11c} on using delayed CSIT to retrospectively compensate for interference due to lack of current CSIT, on the ideas in \cite{KYGY:12o} and later in \cite{YKGY:12d,GJ:12o} on exploiting perfect delayed and imperfect current CSIT, as well as the work in \cite{CE:12d,CE:12c} which - in the context of imperfect and delayed CSIT - introduced encoding and decoding with a phase-Markov structure that will be used later on.  The work here is also motivated by the work in \cite{VV:11t} which considered the use of delayed feedback in different MIMO BC settings, as well as by recent progress in \cite{YYGK:12} that considered MIMO BC and MIMO IC settings that enjoyed perfect delayed feedback as well as imperfect current feedback of a quality that remained unchanged throughout the communication process ($\alpha^{(1)}=  -\lim_{P\rightarrow \infty}  \frac{  \E[||\Ha-\hHa[t]||_{F}^2]  }{  \log P }, \alpha^{(2)}= -\lim_{P\rightarrow \infty}  \frac{  \E[||\Hb-\hHb[t]||_{F}^2]  }{  \log P }, \ \forall t$).  The work is finally motivated by the recent approach in \cite{CE:13it} that employed sequences of evolving quality exponents to address a more fundamental problem of deriving the performance limits given a general CSIT process of a certain quality.


\section{DoF region of the MIMO BC and MIMO IC\label{sec:bc-dof}}

We proceed with the main DoF results, which are proved in Section~\ref{sec:outerb} that describes the outer bound, and in Section~\ref{sec:schemes} that describes an inner bound by extending the schemes from \cite{CE:13it} to the symmetric MIMO BC and MIMO IC cases of interest.
We recall that we consider communication of large duration $n$, a possibly correlated channel process $\{\Ha,\Hb\}_{t=1}^n$ ($\{\Haa,\Hab,\Hba,\Hbb\}_{t=1}^n$ for the IC), and a feedback process of quality defined by the statistics of $\{(\Ha - \hHa),(\Hb - \hHb)\}_{t=1,t'=1}^n$ ($\{(\Hij-\hHij)\}_{t,t' = 1}^n, \ \ i,j=1,2$ for the IC). We henceforth, without loss of generality, label the users so that $\bar{\alpha}^{(2)}\leq \bar{\alpha}^{(1)}$.

We proceed with the DoF region for any CSIT process with sufficiently good delayed CSIT.

\vspace{3pt}
\begin{theorem} \label{thm:IdelayCSIT}
The optimal DoF region of the two-user ($M\times (N,N)$) MIMO BC with a CSIT process $\{\hHa, \hHb\}_{t=1,t'=1}^n$ of quality $\{(\Ha - \hHa),(\Hb - \hHb)\}_{t=1,t'=1}^n$ that has sufficiently good delayed CSIT, is given by
  \begin{align}
	   d_1 &\le \min\{M,N\} \\
		  d_2 &\le \min\{M,N\}   \\
  		 d_1+ d_2 &\le \min\{M, 2N\}   \label{eq:thmBCL0}\\
   \frac{d_1}{\min\{M,N\}}  + \frac{d_2}{\min\{M,2N\}} & \le 1 + \frac{\min\{M,2N \} - \min\{M,N\} }{\min\{M,2N\} } \ \bar{\alpha}^{(1)} \label{eq:thmBCL2}\\
     \frac{d_1}{\min\{M,2N\}}  + \frac{d_2}{\min\{M,N\}} & \le 1 + \frac{\min\{M,2N \} - \min\{M,N\} }{\min\{M,2N\} } \ \bar{\alpha}^{(2)} \label{eq:thmBCL1}
  \end{align}
while for the $(M,M)\times (N,N)$ MIMO IC with feedback quality $\{(\Hij-\hHij)\}_{t,t' = 1}^n, \ \ i,j=1,2$, the above holds after substituting \eqref{eq:thmBCL0} with
\beq \label{eq:ICuniqueBound}
d_1+ d_2 \le \min\{2M, 2N, \max\{M,N\}\}.
\eeq
\end{theorem}
\vspace{3pt}

\begin{figure}
\centering
\includegraphics[width=13cm]{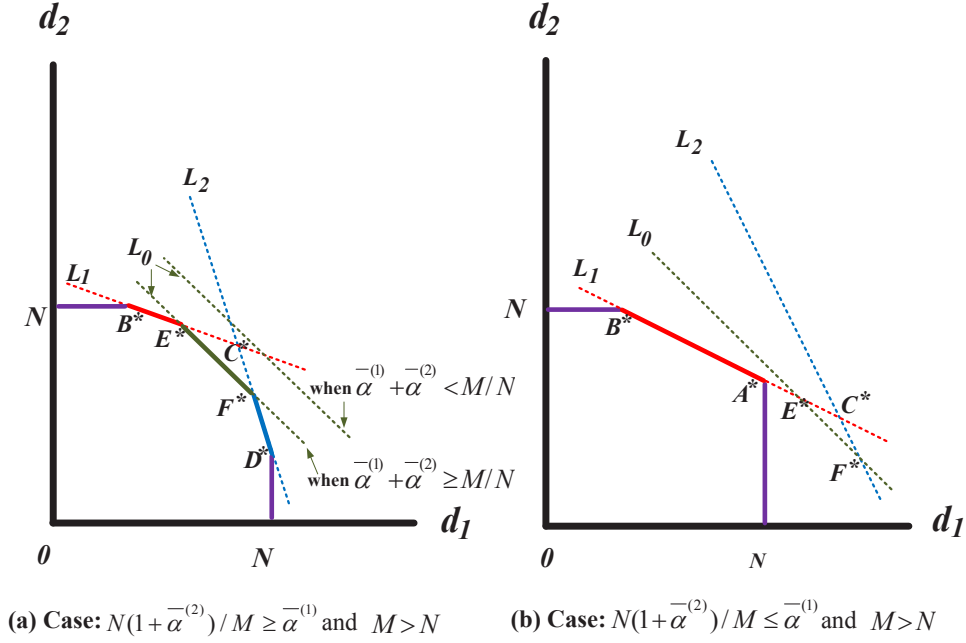}
\caption{Optimal DoF regions for two different cases for the two-user MIMO BC and MIMO IC, with $M>N$ and $\min \{ \bar{\beta}^{(1)}, \bar{\beta}^{(2)}\} \geq \min \{1, M-\min\{M,N\},\frac{ N(1 + \bar{\alpha}^{(1)} + \bar{\alpha}^{(2)}) }{\min\{M,2N\}+N}, \frac{N( 1 +\bar{\alpha}^{(2)}) }{\min\{M,2N\}}  \}$.
The corner points take the following values: $A^{*}=\bigl(N, \frac{(M-N)N( 1 + \bar{\alpha}^{(2)})}{M}\bigr)$, $B^{*}=\bigl((M-N)\bar{\alpha}^{(2)}, N\bigr)$, $C^{*}=\bigl(\frac{MN}{M+N}(1+\bar{\alpha}^{(1)}-\frac{N}{M}\bar{\alpha}^{(2)}), \frac{MN}{M+N}(1+\bar{\alpha}^{(2)}-\frac{N}{M}\bar{\alpha}^{(1)})\bigr)$, $D^{*}=\bigl(N, (M-N)\bar{\alpha}^{(1)}\bigr)$, $E^{*} = \bigl(M-N\bar{\alpha}^{(2)}, \  N\bar{\alpha}^{(2)} \bigr)$, $F^{*}= \big(N\bar{\alpha}^{(1)}, \ M-N\bar{\alpha}^{(1)}\bigr)$. Line~$L_0$ corresponds to the bound in~\eqref{eq:thmBCL0}, Line~$L_1$ corresponds to the bound in~\eqref{eq:thmBCL1}, while line~$L_2$ corresponds to the bound in~\eqref{eq:thmBCL2}.}
\label{fig:MIMODoFAsymmeticCSITcase1andcase2}
\end{figure}

The following proposition provides the DoF region inner bound for the regime of low-quality delayed CSIT. The proof is shown in Section~\ref{sec:schemes}.

\vspace{3pt}
\begin{proposition} \label{prop:MIMOgenCSITInner}
The DoF region of the two-user ($M\times (N,N)$) MIMO BC with a CSIT process of quality $\{(\Ha - \hHa),(\Hb - \hHb)\}_{t=1,t'=1}^n$ such that $\min \{ \bar{\beta}^{(1)}, \bar{\beta}^{(2)}\} < \min \{1, M-\min\{M,N\},\frac{ N(1 + \bar{\alpha}^{(1)} +  \bar{\alpha}^{(2)}) }{\min\{M,2N\}+N}, \frac{N( 1 +\bar{\alpha}^{(2)}) }{\min\{M,2N\}}  \}$, is inner bounded by the polygon described by
  \begin{align}
	   d_1 &\le \min\{M,N\} \\
		  d_2 &\le \min\{M,N\}   \\
  		 d_1+ d_2 &\le \min\{M, 2N\}  \label{eq:BCuniqueInner} \\
	   d_1 +d_2 &\leq  \min\{M,N\} + (\min\{M,2N\}-\min\{M,N\})\min \{ \bar{\beta}^{(1)}, \bar{\beta}^{(2)}\}\\
   \frac{d_1}{\min\{M,N\}}  + \frac{d_2}{\min\{M,2N\}} & \le 1 + \frac{\min\{M,2N \} - \min\{M,N\} }{\min\{M,2N\} } \ \bar{\alpha}^{(1)} \\
     \frac{d_1}{\min\{M,2N\}}  + \frac{d_2}{\min\{M,N\}} & \le 1 + \frac{\min\{M,2N \} - \min\{M,N\} }{\min\{M,2N\} } \ \bar{\alpha}^{(2)}.
  \end{align}
while for the $(M,M)\times (N,N)$ MIMO IC with feedback quality $\{(\Hij-\hHij)\}_{t,t' = 1}^n, \ \ i,j=1,2$, the above holds after substituting \eqref{eq:BCuniqueInner} with
\beq \label{eq:ICuniqueBound2}
d_1+ d_2 \le \min\{2M, 2N, \max\{M,N\}\}.
\eeq
\end{proposition}

\vspace{3pt}

\begin{remark}
As a small comment, and to place the above proposition in the context of previous work, we briefly note that deriving the DoF for even the simplest instance in the setting of low-quality delayed CSIT - corresponding to the case of $\beta^{(1)} = \beta^{(2)} =\alpha^{(1)} = \alpha^{(2)} = 0$ - has been a long lasting open problem. In this simple setting of all-zero quality exponents, the conjectured DoF of $d_1 = d_2 = 1/2$ in \cite{LSW:05}, matches the above inner bound.
\end{remark}

\subsection{Imperfect current CSIT can be as useful as perfect current CSIT}

The above results allow for direct conclusions on the amount of CSIT that is necessary to achieve the optimal DoF performance associated to perfect and immediately available CSIT.  The following corollary holds for the BC and the IC case, for which we also remember that there is no need for CSIT when $M\leq N$ (cf.~\cite{HJSV:12,VV:09}).  
The proofs for the following corollary, and of the corollary immediately after that, are direct from the above theorems.

\vspace{3pt}
\begin{corollary} \label{thm:IcurrentCSIT}
Having a CSIT process that offers $\bar{\alpha}^{(1)}+\bar{\alpha}^{(2)} \geq \min\{M, 2N\}/N $, allows for the optimal sum-DoF associated to having perfect and immediately (full) CSIT ($\bar{\alpha}^{(1)} = \bar{\alpha}^{(2)} = 1$).
\end{corollary}
\vspace{3pt}
The above suggests a reduction in the required feedback quality $\bar{\alpha}^{(1)},\bar{\alpha}^{(2)}$, as the number of receive antennas increases.  Furthermore as stated before, when applied to the IC case, the above also reveals that no CSIT is needed for the direct links.

Along the same lines, the following describes the amount of delayed CSIT that suffices to achieve the DoF associated to perfect delayed CSIT.

\vspace{3pt}
\begin{corollary} \label{thm:IdelayedCSIT}
Any CSIT process that offers $\min \{ \bar{\beta}^{(1)}, \bar{\beta}^{(2)}\} \geq \min \{1, M-\min\{M,N\},\frac{ N(1 + \bar{\alpha}^{(1)} +  \bar{\alpha}^{(2)}) }{\min\{M,2N\}+N}, \frac{N( 1 +\bar{\alpha}^{(2)}) }{\min\{M,2N\}}  \}$, can achieve the same DoF region as a CSIT process that offers perfect delayed CSIT ($\bar{\beta}^{(1)} = \bar{\beta}^{(2)} = 1$).
\end{corollary}
\vspace{3pt}


\section{Outer bound for the MIMO BC and MIMO IC with evolving feedback\label{sec:outerb}}

We proceed to first describe the outer bound for the BC case. The bound, presented in the following lemma, draws from \cite{CYE:13isit} and \cite{CE:13it}, and for this we here mainly focus on the proof steps that are important in the MIMO case.

\vspace{3pt}
\begin{lemma} \label{lm:bc-evol-outerb}
The DoF region of the two-user MIMO BC with a CSIT process $\{\hHa, \hHb\}_{t=1,t'=1}^n$ of quality $(\Ha - \hHa),(\Hb - \hHb)\}_{t=1,t'=1}^n$, is upper bounded as
  \begin{align}
	      d_1 &\le \min\{M,N\} \label{eq:bcupperbound1}\\
		  d_2 &\le \min\{M,N\}   \label{eq:bcupperbound2}\\
		 d_1+ d_2 &\le \min\{M, 2N\}   \label{eq:bcupperbound0}\\
     \frac{d_1}{\min\{M,N\}}  + \frac{d_2}{\min\{M,2N\}} & \le 1 + \frac{\min\{M,2N \} - \min\{M,N\} }{\min\{M,2N\} } \ \bar{\alpha}^{(1)}  \label{eq:bcupperbound3}\\
     \frac{d_1}{\min\{M,2N\}}  + \frac{d_2}{\min\{M,N\}} & \le 1 + \frac{\min\{M,2N \} - \min\{M,N\} }{\min\{M,2N\} } \ \bar{\alpha}^{(2)}. \label{eq:bcupperbound4}
  \end{align}
\end{lemma}
\vspace{3pt}

\begin{proof}
For notational convenience we define $\left\langle \bullet \right\rangle' \defeq \min\left\{ \bullet, M \right\}$, $\Omega_{[n]} \defeq \{ \Ha, \Hb, \hHa, \hHb,\}_{t=1 \ t'=1}^{n \ \ \ n}$, $ \ya[{[n]}] \defeq \{\ya\}_{t=1}^{n}$ and $\quad \yb[{[n]}] \defeq  \{\yb\}_{t=1}^{n}$.

We first design a degraded version of the BC by giving the observations and messages of receiver $1$ to receiver
$2$.  This allows for \begin{align}
  n R_1 &\le I(W_1; \ya[{[n]}] | \Omega_{[n]}) + n \epsilon \\
  n R_2 &\le I(W_2; \ya[{[n]}],\yb[{[n]}] | W_1, \Omega_{[n]}) + n \epsilon
\end{align}
due to Fano's inequality, due to the basic chain-rule of mutual
information, and due to the fact that messages from different users are independent.
This now gives that
\begin{align}
  n R_1 &\le h(\ya[{[n]}] \cond \Omega_{[n]}) - h(\ya[{[n]}] \cond W_1, \Omega_{[n]}) + n \epsilon \\
  n R_2 &\le h(\ya[{[n]}], \yb[{[n]}] \cond W_1, \Omega_{[n]}) - h(\ya[{[n]}], \yb[{[n]}] \cond
  W_1, W_2, \Omega_{[n]}) + n \epsilon
\end{align}
and that
\begin{align}
  &   \frac{nR_1}{\left\langle N\right\rangle'} +\frac{nR_2}{\left\langle 2N\right\rangle'} -\bigl(\frac{n}{\left\langle N\right\rangle'} +\frac{n}{\left\langle 2N\right\rangle'} \bigr)\epsilon  \nonumber \\
  & \le   \frac{1}{\left\langle N\right\rangle'} h(\ya[{[n]}] \cond \Omega_{[n]}) +  \frac{1}{\left\langle 2N\right\rangle'} h(\ya[{[n]}], \yb[{[n]}] \cond W_1, \Omega_{[n]}) -  \frac{1}{\left\langle N\right\rangle'}h(\ya[{[n]}] \cond W_1, \Omega_{[n]}) - \frac{1}{\left\langle 2N\right\rangle'}h(\ya[{[n]}], \yb[{[n]}] \cond
  W_1, W_2, \Omega_{[n]})  \nonumber \\
  & =   \frac{1}{\left\langle N\right\rangle'} h(\ya[{[n]}] \cond \Omega_{[n]}) +  \frac{1}{\left\langle 2N\right\rangle'} h(\ya[{[n]}], \yb[{[n]}] \cond W_1, \Omega_{[n]}) -  \frac{1}{\left\langle N\right\rangle'}h(\ya[{[n]}] \cond W_1, \Omega_{[n]}) + no(\log P) \label{eq:tmp91} \\
 & \leq   n \log P + n o(\log P) +  \frac{1}{\left\langle 2N\right\rangle'} h(\ya[{[n]}], \yb[{[n]}] \cond W_1, \Omega_{[n]}) -  \frac{1}{\left\langle N\right\rangle'}h(\ya[{[n]}] \cond W_1, \Omega_{[n]}) + n o(\log P)      \label{eq:tmp41}  \\
 & =  \sum_{t=1}^n \left( \frac{1}{\left\langle 2N\right\rangle'}  h(\ya[t], \yb[t] \cond  \ya[{[t-1]}], \yb[{[t-1]}], W_1, \Omega_{[n]}) -  \frac{1}{\left\langle N\right\rangle'}h( \ya[t] \cond \ya[{[t-1]}], W_1, \Omega_{[n]}) \right) + n \log P + n o(\log P)\label{eq:tmp61} \\
 & \leq  \sum_{t=1}^n \left( \frac{1}{\left\langle 2N\right\rangle'}  h(\ya[t], \yb[t] \cond  \ya[{[t-1]}], \yb[{[t-1]}], W_1, \Omega_{[n]}) -  \frac{1}{\left\langle N\right\rangle'}h( \ya[t] \cond \ya[{[t-1]}], \yb[{[t-1]}],  W_1, \Omega_{[n]}) \right) + n \log P + n o(\log P) \label{eq:tmp71} \\
 & \leq   \frac{1}{\left\langle 2N\right\rangle' \left\langle N\right\rangle'} \sum_{t=1}^n   \left(  (\left\langle 2N\right\rangle' -\left\langle N\right\rangle' ) \left\langle N\right\rangle' \alpha^{(1)}_t \log P   + o(\log P)  \right) + n \log P + n o(\log P) \label{eq:tmp68}  \\
 & =   \frac{n}{\left\langle 2N\right\rangle' \left\langle N\right\rangle'}   \left(  (\left\langle 2N\right\rangle' -\left\langle N\right\rangle' ) \left\langle N\right\rangle' \bar{\alpha}^{(1)}\log P   +  o(\log P) \right) + n \log P + n o(\log P) \nonumber \\
& =   \frac{n (\left\langle 2N\right\rangle' -\left\langle N\right\rangle' )}{\left\langle 2N\right\rangle' }  \bar{\alpha}^{(1)}\log P  + n \log P + n o(\log P).
\end{align}%
In the above, \eqref{eq:tmp91} is due to the fact that knowledge of $\{W_1, W_2, \Omega_{[n]}\}$ allows for reconstruction of $\ya[{[n]}], \yb[{[n]}]$ up to noise level, while \eqref{eq:tmp41} is due to the fact that $h(\ya[{[n]}] \cond \Omega_{[n]})\le  \left\langle N\right\rangle'\log P + o(\log P)$. Additionally \eqref{eq:tmp61} is due to the chain rule of differential
entropy, \eqref{eq:tmp71} is due to the fact that conditioning reduces differential
entropy, and \eqref{eq:tmp68} is directly from \cite[Proposition~4]{CYE:13isit} after setting
$U = \{\ya[{[t-1]}], \yb[{[t-1]}], W_1, \Omega_{[n]}\}
\setminus \{\Ha, \Hb, \hHa[t], \hHb[t] \}$ \footnote{We note that the result in \cite[Proposition~4]{CYE:13isit} holds for a large family of channel models, under the assumption that the CSIT estimates up to time $t$ are independent of the current estimate errors at time $t$.}.

The above gives the bound in \eqref{eq:bcupperbound3}. Interchanging the roles of the users gives the bound in~\eqref{eq:bcupperbound4}. The bounds in \eqref{eq:bcupperbound1},\eqref{eq:bcupperbound2} are basic single-user constraints, while the bound in \eqref{eq:bcupperbound0} corresponds to an assumption of user cooperation.
\end{proof}

\subsection{Outer bound proof for the IC \label{sec:outerb4IC}}

The task is to show that the above bounds \eqref{eq:bcupperbound1},\eqref{eq:bcupperbound2},\eqref{eq:bcupperbound3},\eqref{eq:bcupperbound4} hold for the case of the IC.  First let us set $\ya, \yb$ to take the form in \eqref{eq:ICmodely1},\eqref{eq:ICmodely2}, and let us denote $\Omega_{[n]} \defeq \{ \Haa, \Hab, \Hbb, \Hba, \hHab, \hHba,\hHaa, \hHbb\}_{t=1 \ t'=1}^{n \ \ \ n} $.  Focusing on the bound in \eqref{eq:bcupperbound3}, we note that the transition from \eqref{eq:tmp71} to \eqref{eq:tmp68} holds in the IC setting, because knowledge of $\{W_1, \Omega_{[n]}\}$ implies knowledge of $\Haa \xa$ and $\Hba \xa$, which in turn implies that
\begin{align*}
&\frac{h(\ya[t], \yb[t] \cond \ya[{[t-1]}], \yb[{[t-1]}], W_1, \Omega_{[n]})}{\left\langle 2N\right\rangle'} - \frac{h( \ya[t] \cond \ya[{[t-1]}], \yb[{[t-1]}], W_1, \Omega_{[n]})}{\left\langle N\right\rangle'} \\
&= \frac{h( \Hab \xb + \za, \Hbb \xb + \zb \cond \ya[{[t-1]}], \yb[{[t-1]}], W_1, \Omega_{[n]})}{\left\langle 2N\right\rangle'} - \frac{h( \Hab \xb + \za \cond \ya[{[t-1]}], \yb[{[t-1]}], W_1, \Omega_{[n]})}{\left\langle N\right\rangle'}
\end{align*}
which has the form of the difference of the differential entropies corresponding to a BC setting, where the role of the BC transmitter is replaced now by the second IC transmitter. This implies that \eqref{eq:bcupperbound3} holds for the IC.  Bounds \eqref{eq:bcupperbound1},\eqref{eq:bcupperbound2},\eqref{eq:bcupperbound4} follow easily. Finally, the bound $d_1 +d_2 \leq \min\{2M, 2N, \max\{M,N\},\max\{M,N\}\}$ is directly from \cite{JF:07}.


\section{Phase-Markov transceiver for imperfect and delayed feedback\label{sec:schemes}}

We proceed to extend the MISO BC scheme in \cite{CE:13it}, to the current setting of the MIMO BC and MIMO IC.

While part of the extension of the schemes in \cite{CE:13it} involves keeping track of the dimensionality changes that come with MIMO, there are here modifications that are not trivial.  These include changes in the way the scheme performs interference quantization as well as power and rate allocation, differences in decoding, as well as differences in the way the information is aggregated to achieve the corresponding DoF corner points. A particular extra challenge corresponding to the MIMO IC, has to do with the fact that now the signals must be sent by two independent transmitters. This will reflect on the power and rate allocation at each transmitter, and on the way common and private information is decoded at each receiver.

Before proceeding with the schemes, we again note that we only need to consider the case where $N< M \leq 2N$ simply because, both for the BC and the IC, the optimal DoF can be achieved without any CSIT whenever $M \leq N$, while having $M>2N$ can be shown to be equivalent, in terms of DoF, with the case of having $M = 2N$.
We first begin with the scheme description for the BC setting, while at the end we will describe the modifications required to achieve the result for the IC case.
Section~\ref{sec:scheme_en} will describe the encoding part, Section~\ref{sec:scheme_de} the decoding part, and Section~\ref{sec:scheme_DoFa} will describe how we calibrate the parameters of this universal scheme to achieve the different DoF corner points.

The challenge here will be to design a scheme of large duration $n$, that utilizes the CSIT process $\{\hHa,\hHb\}_{t=1,t'=1}^n$.  As in~\cite{CE:13it}, the causal scheme will not require knowledge of future quality exponents, nor of predicted CSIT estimates of future channels. We remind the reader that the users are labeled so that $\bar{\alpha}^{(2)}\leq \bar{\alpha}^{(1)}$. 

For notational convenience, we will use
\beq \hHta \defeq \hHa[t],  \quad \hHtb \defeq \hHb[t]\eeq
\beq \cHta \defeq \hHa[t+\eta] ,  \quad \cHtb \defeq \hHb[t+\eta] \eeq
to denote the current and delayed estimates of $\Ha,\Hb$, with the corresponding estimation errors being
\begin{equation}
\label{eq:MMSEc}
  \tHta \defeq \Ha - \hHta , \quad     \tHtb \defeq \Hb - \hHtb
\end{equation}
\begin{equation}
\label{eq:MMSEd}
  \dHta \defeq \Ha - \cHta , \quad     \dHtb \defeq \Hb - \cHtb.
\end{equation}

We will also use the notation
\beq \label{eq:NotationPower}
P^{(\ev)}_t \defeq \E |\ev_t|^2
\eeq to denote the power of a symbol $\ev_t$ corresponding to time-slot~$t$, and we will use $r^{(\ev)}_t$ to denote the prelog factor of the number of bits
$r^{(\ev)}_t\log P - o(\log P)$ carried by symbol $\ev_t$ at time $t$.

\subsection{Encoding \label{sec:scheme_en}}
As in \cite{CE:13it}, we subdivide the overall time duration $n$, into $S$ phases, each of duration of $T$, such that each phase~$s$ ($s = 1,2,\cdots,S$) takes place over the time slots $t\in\Bc_s$
\begin{align}
\Bc_s = \{\Bc_{s,\ell}  \defeq (s\!-\!1)2T +\ell  \}_{\ell=1}^{T}, \quad   s=1,\cdots,S. \label{eq:timephase}
\end{align}
Naturally in the gap of what we define here to be consecutive phases, another message is sent, using the same exact scheme.
Going back to the aforementioned assumption, $T$ is sufficiently large so that
\begin{align} \label{eq:average}
\frac{1}{T}\sum_{t\in \Bc_s} \alpha^{(i)}_t \!\rightarrow\!  \bar{\alpha}^{(i)} , \  \frac{1}{T}\sum_{t\in \Bc_s}  \beta^{(i)}_t \!\rightarrow\!  \bar{\beta}^{(i)},  \ s=1,\cdots,S
\end{align}
$i=1,2$.  For notational convenience we will also assume that $T>\eta$ (cf.~\eqref{eq:alpha1}), although this assumption can be readily removed, as this was argued in \cite{CE:13it}. Finally with $n$ being infinite, $S$ is also infinite.

Adhering to a phase-Markov structure which - in the context of imperfect and delayed CSIT, was first introduced in~\cite{CE:12d,CE:12c} - the scheme will quantize the accumulated interference of a certain phase~$s$, broadcast it to both receivers over phase~$(s+1)$, while at the same time it will send extra information to both receivers in phase~$s$, which will help recover the interference accumulated in phase~$(s-1)$.

We first describe the encoding for all phases except the last phase which will be addressed separately due to its different structure.

\subsubsection{Phase~$s$,  for $s=1,2,\cdots,S-1$}
In each phase, the scheme combines zero forcing and superposition coding, power and rate allocation, and interference quantizing and broadcasting. We proceed to describe these steps.

\paragraph{Zero forcing and superposition coding} At time $t\in \Bc_s$ (of phase $s$), the transmitter sends
\begin{align} \label{eq:TxX1Ph1}
\xv_{t} =  \Wm_t  \cv_t + \Um_t \av_t + \Um_t^{'} \av^{'}_t + \Vm_t \bv_t + \Vm_t^{'} \bv^{'}_t \end{align}
where $\av_t \in  \CC^{ (M-N)\times 1} , \av^{'}_t \in  \CC^{ N \times 1 }$ are the vectors of symbols meant for receiver~1, $\bv_t \in \CC^{ (M-N)\times 1} , \bv^{'}_t \in \CC^{ N\times 1} $ are those meant for receiver~2, where $\cv_t \in  \CC^{ M\times M} $ is a common symbol vector, where $\Um_t= (\hHtb)^{\bot} \in \CC^{ M \times (M-N)}$ is a unit-norm matrix that is orthogonal to $\hHtb$, where $ \Vm_t= (\hHta)^{\bot} \in \CC^{ M \times (M-N)}$ is orthogonal to $\hHta$, and where $\Wm_t \in \CC^{ M \times M}, \Um_t^{'}  \in \CC^{ M \times N}, \Vm_t^{'}   \in \CC^{ M \times N} $ are predetermined randomly-generated matrices known by all nodes.

\paragraph{Power and rate allocation}
The powers and (normalized) rates during phase $s$ time-slot $t$, are
\begin{equation}\label{eq:RPowerX1Ph1}
\begin{array}{lllll}
P^{(\cv)}_t \doteq  P , &  P^{(\av)}_t \doteq  P^{\delta_t^{(2)}}, & P^{(\bv)}_t \doteq  P^{\delta_t^{(1)}}, & P^{(\av')}_t \doteq P^{\delta_t^{(2)}-\alpha^{(2)}_t}, & P^{(\bv')}_t \doteq P^{\delta_t^{(1)}-\alpha^{(1)}_t}\\
& r^{(\av)}_t  = (M-N)\delta_t^{(2)} , & r^{(\bv)}_t = (M-N)\delta_t^{(1)} , &  r^{(\av')}_t =N(\delta_t^{(2)}-\alpha^{(2)}_t)^{+} , & r^{(\bv')}_t =N(\delta_t^{(1)}-\alpha^{(1)}_t)^{+}. \end{array}
\end{equation}
where $\{\delta_t^{(1)}, \delta_t^{(2)}\}_{t\in\Bc_s}$ are designed such that
\begin{align}
\beta_t^{(i)} \geq \delta_t^{(i)}  \quad & i=1,2,  \ t\in\Bc_s   \label{eq:powerde}\\
\frac{1}{T}\sum_{t\in \Bc_s}\delta_t^{(1)}  & =   \frac{1}{T}\sum_{t\in \Bc_s}\delta_t^{(2)} = \bar{\delta}  \label{eq:powerde1} \\
\frac{1}{T}\sum_{t\in \Bc_s} (\delta_t^{(i)}- \alpha_t^{(i)})^{+}  & =(  \bar{\delta} - \bar{\alpha}^{(i)})^{+} \quad i=1,2 \label{eq:powerde2}
\end{align}
for some $\bar{\delta}$ that will be bounded by
\begin{align} \label{eq:bardelta}
\bar{\delta} \leq \min \{ 1,\bar{\beta}^{(1)},  \bar{\beta}^{(2)}, \frac{ N(1 + \bar{\alpha}^{(1)} +  \bar{\alpha}^{(2)}) }{M+N}, \frac{N( 1 +\bar{\alpha}^{(2)}) }{M}  \}
\end{align}
and which will be set to specific values later on, depending on the DoF corner point we wish to achieve.

The exact solutions for $\{\delta_t^{(1)}, \delta_t^{(2)}\}_{t\in\Bc_s}$  satisfied \eqref{eq:powerde},\eqref{eq:powerde1},\eqref{eq:powerde2} are shown in \cite{CE:13it}, and the rates of the common symbols
$\{\cv_{\Bc_{s,t} }\}_{t=1}^T$ are designed to jointly carry
\begin{align}
 T(N-(M-N)\bar{\delta}) \log P -o(\log P) \label{eq:ratejointc}
\end{align}
bits.

To put the above allocation in perspective, we show the received signals, and describe under each term the order of the summand's average power.  These signals take the form
\begin{align}
  \ya &= \underbrace{ \Ha \Wm_t \cv_t}_{P}+\underbrace{\Ha \Um_t \av_t}_{P^{\delta_t^{(2)}}} +\underbrace{\Ha \Um_t^{'} \av_t^{'}}_{P^{\delta_t^{(2)}-\alpha^{(2)}_t}} +\underbrace{\za}_{P^0} +\overbrace{\underbrace{\cHta ( \Vm_t \bv_t+ \Vm_t^{'} \bv_t^{'})}_{P^{\delta_t^{(1)}-\alpha^{(1)}_t}}}^{\check{\iota}^{(1)}_t} +  \overbrace{\underbrace{\dHta ( \Vm_t \bv_t+ \Vm_t^{'} \bv_t^{'})}_{P^{\delta_t^{(1)}-\beta^{(1)}_t}\leq P^{0}}}^{\iota^{(1)}_t-\check{\iota}^{(1)}_t} \label{eq:sch1y1}\\
  \yb	&= \underbrace{\Hb \Wm_t \cv_t}_{P}+\underbrace{\Hb \Vm_t \bv_t}_{P^{\delta_t^{(1)}}}+\underbrace{\Hb \Vm_t^{'} \bv_t^{'}}_{P^{\delta_t^{(1)}-\alpha^{(1)}_t}}+\underbrace{\zb}_{P^0}  +\overbrace{\underbrace{\cHtb ( \Um_t \av_t+ \Um_t^{'} \av_t^{'})}_{P^{\delta_t^{(2)}-\alpha^{(2)}_t}}}^{\check{\iota}^{(2)}_t} + \overbrace{\underbrace{\dHtb ( \Um_t \av_t+ \Um_t^{'} \av_t^{'})}_{ P^{\delta_t^{(2)}-\beta^{(2)}_t}\leq P^{0}}}^{\iota^{(2)}_t-\check{\iota}^{(2)}_t} \label{eq:sch1y2}
\end{align}
where
\beq \label{eq:c2barsg}
\iota^{(1)}_t  \defeq \Ha(\Vm_t \bv_t + \Vm_t^{'} \bv_t^{'}), \
\iota^{(2)}_t  \defeq \Hb (\Um_t \av_t + \Um_t^{'} \av_t^{'})
\eeq
denote the interference at receiver~1 and receiver~2 respectively, and where
\beq \label{eq:cbarg}
\check{\iota}^{(1)}_t  \defeq \cHta(\Vm_t \bv_t+\Vm_t^{'} \bv_t^{'}) , \
\check{\iota}^{(2)}_t  \defeq \cHtb(\Um_t \av_t + \Um_t^{'} \av_t^{'})
\eeq
denote the transmitter's delayed estimates of $\iota^{(1)}_t,\iota^{(2)}_t$.

\paragraph{Quantizing and broadcasting the accumulated interference}  Before the beginning of phase~$(s+1)$, the transmitter reconstructs $\check{\iota}^{(1)}_t, \check{\iota}^{(2)}_t $ for all $t\in \Bc_s$, using its knowledge of delayed CSIT, and quantizes these into
\beq\label{eq:quntisch1}
  \bar{\check{\iota}}^{(1)}_t = \check{\iota}^{(1)}_t -\tilde{\iota}^{(1)}_t, \quad
	\bar{\check{\iota}}^{(2)}_t = \check{\iota}^{(2)}_t - \tilde{\iota}^{(2)}_t
\eeq
using a total of $N(\delta_t^{(1)}-\alpha^{(1)}_t)^{+}\log P$ and $N(\delta_t^{(2)}-\alpha^{(2)}_t)^{+}\log P$ quantization bits respectively.  This allows for bounded power of quantization noise $\tilde{\iota}^{(1)}_t, \tilde{\iota}^{(2)}_t$, i.e, allows for $\E|\tilde{\iota}^{(2)}_t|^2 \doteq \E|\tilde{\iota}^{(1)}_t|^2 \doteq 1$, since $\E|\check{\iota}^{(2)}_t|^2 \doteq P^{\delta^{(2)}_t-\alpha^{(2)}_t}, \ \E|\check{\iota}^{(1)}_t|^2 \doteq P^{\delta^{(1)}_t-\alpha^{(1)}_t}$ (cf.~\cite{CT:06}).
Then the transmitter evenly splits the
\beq\label{eq:bitquan}
\sum_{t\in \Bc_s} \left(  N(\delta_t^{(1)}-\alpha^{(1)}_t)^{+} + N(\delta_t^{(2)}-\alpha^{(2)}_t)^{+} \right)\log P = T N\left(  (\bar{\delta}-\bar{\alpha}^{(1)})^{+} + (\bar{\delta}-\bar{\alpha}^{(2)})^{+} \right)\log P
\eeq
(cf. \eqref{eq:powerde2}) quantization bits into the common symbols $\{\cv_t\}_{t\in\Bc_{s+1}}$ that will be transmitted during the next phase (phase~$s+1$), and which will convey these quantization bits together with other new information bits for the receivers. These $\{\cv_t\}_{t\in\Bc_{s+1}}$ will help the receivers cancel interference, as well as will serve as extra observations (see~\eqref{eq:firstMIMOx1} later on) that will allow for decoding of all private information (see~Table~\ref{tab:bitsummary}).

\begin{table}
\caption{Number of bits carried by private and common symbols, and by the quantized interference (phase~$s$).}
\begin{center}
\begin{tabular}{|c|c|}
  \hline
                & Total bits ($\times \log P$)\\
   \hline
   Private symbols for user~1    & $T((M-N)\bar{\delta}+N(\bar{\delta}- \bar{\alpha}^{(2)})^{+}  )$ \\
   \hline
   Private symbols for user~2    & $T((M-N)\bar{\delta}+N(\bar{\delta}- \bar{\alpha}^{(1)})^{+}  )$ \\
   \hline
   Common symbols        &$T(N-(M-N)\bar{\delta})$  \\
    \hline
  Quantized interference       &$TN((\bar{\delta}-\bar{\alpha}^{(1)})^{+} + (\bar{\delta}-\bar{\alpha}^{(2)})^{+})$  \\
    \hline
\end{tabular}
\end{center}
\label{tab:bitsummary}
\end{table}

Finally, for the last phase~$S$, the main target will be to recover the information on the interference accumulated in phase~$(S-1)$. For large $S$, this last phase can focus entirely on transmitting common symbols.

This concludes the part of encoding, and we now move to decoding.

\subsection{Decoding \label{sec:scheme_de}}

In accordance to the phase-Markov structure, we consider decoding that moves backwards, from the last to the first phase. The last phase was specifically designed to allow decoding of the common symbols $\{\cv_t\}_{t\in \Bc_{S}}$.  Hence we focus on the rest of the phases, to see how - with knowledge of common symbols from the next phase - we can go back one phase and decode its symbols.

During phase~$s$, each receiver uses $\{\cv_t\}_{t\in \Bc_{s+1}}$ to reconstruct the delayed estimates $\{\bar{\check{\iota}}^{(2)}_t,\bar{\check{\iota}}^{(1)}_t\}_{t\in \Bc_s}$, to remove - up to noise level - all
the interference $\iota^{(i)}_t, \ t\in \Bc_s$, by subtracting the delayed interference estimates $\bar{\check{\iota}}^{(i)}_t$ from $\yv^{(i)}_t$.

Now given $\{\bar{\check{\iota}}^{(2)}_t,\bar{\check{\iota}}^{(1)}_t\}_{t\in \Bc_s}$, receiver~1 combines $\{\bar{\check{\iota}}^{(2)}_t\}_{t\in \Bc_s}$ with $\{\ya - \bar{\check{\iota}}^{(1)}_t \}_{t\in \Bc_s}$ to decode $\{\cv_t, \av_t, \av_t^{'}\}_{t\in \Bc_s}$ of phase~$s$. This is achieved by decoding over al accumulated MIMO multiple-access channel (MIMO MAC) of the general form
\begin{align} \label{eq:firstMIMOx1}
\Bmatrix{ \!\!\ya[{\Bc_{s,1}}]-\bar{\check{\iota}}^{(1)}_{\Bc_{s,1}}
          \\ \bar{\check{\iota}}^{(2)}_{\Bc_{s,1}}
					\\   \vdots
					\\  \ya[{\Bc_{s,T}}]-\bar{\check{\iota}}^{(1)}_{\Bc_{s,T}}
          \\  \bar{\check{\iota}}^{(2)}_{\Bc_{s,T}} \!\!}   &=\Bmatrix{ \!\!\!\! \begin{array}{ccc}
\Ha[{\Bc_{s,1}}]\Wm_{\Bc_{s,1}}\!\!\! &        &   \\
 \mathbf{0}        &        &   \\
                   & \!\!\!\!\!\! \!\!\ddots \!\!\!\!\! \!\!\!\!& \\
                   &        & \!\!\!\Ha[{\Bc_{s,T}}]\Wm_{\Bc_{s,T}} \!\!\\
									 &        &  \mathbf{0}   \end{array} \!\!\!}  \!\!\! \Bmatrix{\! \cv_{\Bc_{s,1}} \\ \vdots \\  \cv_{\Bc_{s,T}}\!}
\nonumber \\   &\quad \quad + \Bmatrix{\!\!  \begin{array}{ccc}
\Bmatrix{ \!\! \Ha[{\Bc_{s,1}}] \\ \cHtb[{\Bc_{s,1}}] } \!\! \Bmatrix{\!  \Um_{\Bc_{s,1}}  \    \Um_{\Bc_{s,1}}^{'} \!\!} &        &   \\
                   & \!\!\!\!\!\!\!\!\!\ddots \!\!\!\!\!\!\!\!\! & \\
									 &        &  \Bmatrix{ \! \Ha[{\Bc_{s,T}}] \\ \cHtb[{\Bc_{s,T}}] \!\!\!} \! \!\Bmatrix{ \Um_{\Bc_{s,T}}  \    \Um_{\Bc_{s,T}}^{'} \!\!}  \end{array} \!\!\!\!} \! \!\Bmatrix{ \av_{\Bc_{s,1}} \\  \av_{\Bc_{s,1}}^{'} \\  \vdots \\  \av_{\Bc_{s,T}} \\  \av_{\Bc_{s,T}}^{'} }
 \!\!+ \!\!\Bmatrix{\! \tilde{\zv}^{(1)}_{\Bc_{s,1}} \\   -\tilde{\iota}^{(2)}_{\Bc_{s,1}}  \\  \vdots \\  \tilde{\zv}^{(1)}_{\Bc_{s,T}} \\  -\tilde{\iota}^{(2)}_{\Bc_{s,T}}  \!}
\end{align}
where \[\tilde{\zv}^{(1)}_t= \dHta (  \Vm_t  \bv_t + \Vm_t^{'} \bv_t^{'})+ \za + \tilde{\iota}^{(1)}_t\] and where $\E|\tilde{\zv}^{(1)}_t|^2 \doteq 1$.
It can be readily shown (cf.~\cite{CT:06}) that optimal decoding in such a MIMO MAC setting, allows user~1 to achieve the aforementioned rates in \eqref{eq:RPowerX1Ph1},\eqref{eq:powerde1},\eqref{eq:powerde2},\eqref{eq:ratejointc} i.e., allows for $r^{*(\av)}\log P = T\bigl((M-N)\bar{\delta} +N (\bar{\delta} -  \bar{\alpha}^{(2)})^{+}  \bigr) \log P$ bits to be reliably carried by $\Bmatrix{\av_{\Bc_{s,1}} \  \av_{\Bc_{s,1}}^{'}  \cdots \  \av_{\Bc_{s,T}} \ \av_{\Bc_{s,T}}^{'} }^\T$,  as well as allows for $r^{*(\cv)} \log P = T\bigl(N-(M-N)\bar{\delta}\bigr) \log P$ bits to be carried by $\Bmatrix{\cv_{\Bc_{s,1}} \ \cdots \  \cv_{\Bc_{s,T}}}^\T$ .
Similarly receiver~2 can again accumulate enough received signals to construct a similar MIMO MAC, which will again allow for decoding of its own private and common symbols at the aforementioned rates in \eqref{eq:RPowerX1Ph1},\eqref{eq:ratejointc}.

Now the decoders shift to phase~$s-1$ and use $\{\cv_t\}_{t\in \Bc_s}$ to decode the common and private symbols of that phase. Decoding stops after decoding of the symbols in phase~1.

\subsection{Calibrating the scheme to achieve DoF corner points\label{sec:scheme_DoFa}}
We now describe how to regulate the scheme's parameters to achieve the different DoF points of interest. As previously discussed, we can focus - without an effect to our result\footnote{We clarify that the outer bound left open the possibility that $M>2N$.} - on the case where $N<M\leq 2N$.

Focusing first on the DoF region of the outer bound in Lemma~\ref{lm:bc-evol-outerb}, we note that the DoF corner points that define this region, vary from case to case, as these cases are each defined by each of the following inequalities
\bea
\min \{ \bar{\beta}^{(1)}, \bar{\beta}^{(2)}\} & \geq & \min \{1, M-\min\{M,N\},\frac{ N(1 + \bar{\alpha}^{(1)} +  \bar{\alpha}^{(2)}) }{\min\{M,2N\}+N}, \frac{N( 1 +\bar{\alpha}^{(2)}) }{\min\{M,2N\}}  \} \label{eq:betaSuf}\\
\min \{ \bar{\beta}^{(1)}, \bar{\beta}^{(2)}\} & < & \min \{1, M-\min\{M,N\},\frac{ N(1 + \bar{\alpha}^{(1)} +  \bar{\alpha}^{(2)}) }{\min\{M,2N\}+N}, \frac{N( 1 +\bar{\alpha}^{(2)}) }{\min\{M,2N\}}  \}\label{eq:betaSufNot}\\
\bar{\alpha}^{(1)}&<&\frac{N(1+\bar{\alpha}^{(2)})}{M}\label{eq:alpha1Cond}\\
\bar{\alpha}^{(1)}&\geq & \frac{N(1+\bar{\alpha}^{(2)})}{M}\label{eq:alpha1CondNot}\\
\bar{\alpha}^{(1)} +  \bar{\alpha}^{(2)} &>& \frac{M}{N}\label{eq:alpha12Cond}\\
\bar{\alpha}^{(1)} +  \bar{\alpha}^{(2)} & \leq & \frac{M}{N}.\label{eq:alpha12CondNot}
\eea
The set of all corner points (see also Figure~\ref{fig:MIMODoFAsymmeticCSITcase1andcase2}) is as follows
\bea
A^{*} &=&\bigl(N, \frac{(M-N)N( 1 + \bar{\alpha}^{(2)})}{M}\bigr) \label{eq:Astar}\\
B^{*} &= & \bigl((M-N)\bar{\alpha}^{(2)}, N\bigr) \label{eq:Bstar}\\
C^{*} &=& \bigl(\frac{MN}{M+N}(1+\bar{\alpha}^{(1)}-\frac{N}{M}\bar{\alpha}^{(2)}), \frac{MN}{M+N}(1+\bar{\alpha}^{(2)}-\frac{N}{M}\bar{\alpha}^{(1)})\bigr) \label{eq:Cstar}\\
D^{*} &=& \bigl(N, (M-N)\bar{\alpha}^{(1)}\bigr) \label{eq:Dstar}\\
E^{*} &=& \bigl(M-N\bar{\alpha}^{(2)}, \  N\bar{\alpha}^{(2)} \bigr) \label{eq:Estar}\\
F^{*} &=& \big(N\bar{\alpha}^{(1)}, \ M-N\bar{\alpha}^{(1)}\bigr)\label{eq:Fstar}.
\eea
To achieve the entirety of the outer bound, we need sufficiently good delayed CSIT, and specifically we need~\eqref{eq:betaSuf} to hold. Given~\eqref{eq:betaSuf}, if~\eqref{eq:alpha1Cond} and \eqref{eq:alpha12Cond} hold then the `active' outer bound corner points are $D^{*},B^{*},E^{*},F^{*}$, whereas if \eqref{eq:alpha1Cond} and \eqref{eq:alpha12CondNot} hold, then the active corner points are $D^{*},B^{*},C^{*}$, while if \eqref{eq:alpha1CondNot} holds then the active outer bound corner points are $B^{*},A^{*}$  (see Table~\ref{tab:outerBoundCornerPoints}).

\begin{table}
\caption{Outer bound corner points.}
\begin{center}
\begin{tabular}{|c|c|}
  \hline
  Cases              & Corner points\\
   \hline
   \eqref{eq:alpha1Cond} and \eqref{eq:alpha12Cond}     & $D^{*},B^{*},E^{*},F^{*}$ \\
   \hline
   \eqref{eq:alpha1Cond} and \eqref{eq:alpha12CondNot}   & $D^{*},B^{*},C^{*}$ \\
   \hline
   \eqref{eq:alpha1CondNot}        & $B^{*},A^{*}$ \\
    \hline
\end{tabular}
\end{center}
\label{tab:outerBoundCornerPoints}
\end{table}

We proceed to show how the designed scheme achieves the above points.  To do so, we will show how the scheme, in its general form, achieves a range of DoF points (see~\eqref{eq:DoFd1},\eqref{eq:DoFd2} later on), which can be shifted to the DoF corner points of interest by properly adapting the power allocation and the rate splitting of the new information carried by the common symbols.

\subsubsection{General DoF point}

Remaining on the case where \eqref{eq:betaSuf} holds, we see that the bound in~\eqref{eq:bardelta} now implies that
\begin{align} \label{eq:bardeltaCase1}
\bar{\delta}  \leq \min \{ 1,\frac{ N(1 + \bar{\alpha}^{(1)} +  \bar{\alpha}^{(2)}) }{M+N}, \frac{N( 1 +\bar{\alpha}^{(2)}) }{M}  \}.
\end{align}
Changing $\deltabar$ - within the bounds of \eqref{eq:bardeltaCase1} - will achieve the different DoF points.  Such changing of $\deltabar$, amounts to changing the power allocation (cf.~\eqref{eq:RPowerX1Ph1}) by changing $\{\delta_t^{(1)}, \delta_t^{(2)}\}_{t\in\Bc_s}$ which are a function of $\deltabar$ (cf.~\eqref{eq:powerde1},\eqref{eq:powerde2}).

The first step is to see that for any fixed $\bar{\delta}$, the rate allocation in~\eqref{eq:RPowerX1Ph1} tells us that, the total amount of information, for user 1, in the \emph{private symbols} of a certain phase $s<S$, is equal to
\beq \label{eq:PrivateAmountUser1}
\bigl((M-N)\bar{\delta} + N( \bar{\delta} -  \bar{\alpha}^{(2)})^{+}\bigr) T \log P \eeq
bits, while for user 2 this is
\beq \label{eq:PrivateAmountUser2}
\bigl((M-N)\bar{\delta} + N( \bar{\delta} -  \bar{\alpha}^{(1)})^{+}\bigr) T \log P \eeq
bits.

The next step is to see how much interference there is to load onto these symbols. Given the power and rate allocation in \eqref{eq:powerde},\eqref{eq:powerde1},\eqref{eq:powerde2},\eqref{eq:bardelta}, it is guaranteed that the accumulated quantized interference in a phase $s<S$ has $\bigl(N( \bar{\delta} - \bar{\alpha}^{(1)})^{+} - N( \bar{\delta} - \bar{\alpha}^{(2)})^{+}\bigr)T \log P$ bits (cf.~\eqref{eq:bitquan}), which `fit' into the common symbols of the next phase $(s+1)$ that can carry a total of $\bigl(N- (M-N)\bar{\delta}\bigr)T \log P$ bits (cf.~\eqref{eq:ratejointc}).  This leaves an extra space of $\Delta_{\text{com}} T \log P$ bits in the common symbols, where
\begin{align}
 \Delta_{\text{com}} &\defeq  \bigl( N- (M-N)\bar{\delta} - N( \bar{\delta} - \bar{\alpha}^{(1)})^{+} - N( \bar{\delta} - \bar{\alpha}^{(2)})^{+}  \bigr)   \label{eq:com_inter}
\end{align}
is guaranteed to be non-negative due to \eqref{eq:powerde},\eqref{eq:powerde1},\eqref{eq:powerde2},\eqref{eq:bardelta}. This extra space can be split between the two users, by allocating $\omega \Delta_{\text{com}} T \log P$ bits for the message of user 1, and the remaining $(1-\omega) \Delta_{\text{com}} T \log P$ bits for the message of user~2, for some $\omega\in[0,1]$.

Consequently, considering \eqref{eq:PrivateAmountUser1},\eqref{eq:PrivateAmountUser2}, and given \eqref{eq:betaSuf}, the scheme allows for DoF performance in its general form\footnote{This expression considers that $S$ is large, and thus removes the effect of having a last phase that carries no new message information.}
\begin{align}
d_1 &= (M-N)\bar{\delta} + N( \bar{\delta} -  \bar{\alpha}^{(2)})^{+} +\omega\Delta_{\text{com}}  \label{eq:DoFd1}\\
d_2 &= (M-N)\bar{\delta} + N( \bar{\delta} -  \bar{\alpha}^{(1)})^{+}  +(1-\omega)\Delta_{\text{com}}.  \label{eq:DoFd2}
\end{align}

Again under the setting of \eqref{eq:betaSuf}, we can now move to the different cases, and set $\omega$ and $\deltabar$ (and thus $\Delta_{\text{com}}$) to achieve the different DoF corner points (cf.~Table~\ref{tab:outerBoundCornerPoints}).\\[3pt]
\emph{Case 1 - \eqref{eq:betaSuf} and \eqref{eq:alpha1Cond} and \eqref{eq:alpha12Cond} (points $D^{*},B^{*},E^{*},F^{*}$):}

We first consider the case where \eqref{eq:alpha1Cond} and \eqref{eq:alpha12Cond} hold, and show how to achieve DoF corner points $D^{*},B^{*},E^{*},F^{*}$.
In this setting, \eqref{eq:bardeltaCase1} gives that $\bar{\delta}  \leq 1$ because \eqref{eq:alpha1Cond} implies that $ \bar{\alpha}^{(2)} \leq \bar{\alpha}^{(1)} \leq \frac{ N( 1+\bar{\alpha}^{(1)} +  \bar{\alpha}^{(2)} ) }{M+N} \leq \frac{ N( 1+  \bar{\alpha}^{(2)}) }{M}$ while at the same time \eqref{eq:alpha12Cond} implies that $ \frac{ N(1 + \bar{\alpha}^{(1)} +  \bar{\alpha}^{(2)}) }{M+N} \geq \frac{ N(1 + \frac{M}{N}) }{M+N} =1 $.

To achieve $E^{*} = \bigl(M-N\bar{\alpha}^{(2)}, \  N\bar{\alpha}^{(2)} \bigr)$, we set $\bar{\delta}  =  1,\omega = 0$ (cf.~\eqref{eq:DoFd1}), which gives
\begin{align} d_1 &= (M-N)\bar{\delta} + N( \bar{\delta} -  \bar{\alpha}^{(2)})^{+} \label{eq:d1Gen1}\\
 & =  M  -  N\bar{\alpha}^{(2)} \label{eq:d1Gen2}   \\
d_2 &=  (M-N)\bar{\delta} + N( \bar{\delta} -  \bar{\alpha}^{(1)})^{+} + \Delta_{\text{com}} \label{eq:d2Gen1}\\
 & =  (M -N)  + N( 1 -  \bar{\alpha}^{(1)})^{+} + \bigl( N- (M-N) - N( 1 - \bar{\alpha}^{(1)})^{+} - N( 1 - \bar{\alpha}^{(2)})^{+}  \bigr)  =     N \bar{\alpha}^{(2)}   \label{eq:d2Gen2}
\end{align}
where \eqref{eq:d1Gen1} and \eqref{eq:d2Gen1} are directly from~\eqref{eq:DoFd1},\eqref{eq:DoFd2} after setting $\omega = 0$, and where \eqref{eq:d1Gen2} and \eqref{eq:d2Gen2} consider the value of $\Delta_{\text{com}}$ in \eqref{eq:com_inter} and the fact that $\bar{\delta}  =  1$.

To achieve $F^{*} = \big(N\bar{\alpha}^{(1)}, \ M-N\bar{\alpha}^{(1)}\bigr)$, we set $\bar{\delta}  =  1$ and we set
$\omega = \frac{N( 1 +  \bar{\alpha}^{(1)} +\bar{\alpha}^{(2)})-(M+N) \bar{\delta} }{\Delta_{\text{com}} } = 1$, to get
\begin{align} d_1 &= (M-N)\bar{\delta} + N( \bar{\delta} -  \bar{\alpha}^{(2)})^{+} + \omega\Delta_{\text{com}} \label{eq:Fd1Gen1}\\
  & =  (M -N)  + N( 1 -  \bar{\alpha}^{(2)})^{+} + \Delta_{\text{com}}=    N\bar{\alpha}^{(1)}\label{eq:Fd1Gen2}  \\
d_2 &=  (M-N)\bar{\delta} + N( \bar{\delta} -  \bar{\alpha}^{(1)})^{+} + (1-\omega)\Delta_{\text{com}} \label{eq:Fd2Gen1}\\
& =  (M -N)  + N( 1 -  \bar{\alpha}^{(1)})^{+} =   M-  N \bar{\alpha}^{(1)} \label{eq:Fd2Gen2}
\end{align}
where again \eqref{eq:Fd1Gen1} and \eqref{eq:Fd2Gen1} are directly from~\eqref{eq:DoFd1},\eqref{eq:DoFd2}, and where \eqref{eq:Fd1Gen2} and \eqref{eq:Fd2Gen2} consider the value of $\Delta_{\text{com}}$ in \eqref{eq:com_inter}, together with the fact that $\bar{\delta}  =  1$.

\vspace{1pt}
To achieve $B^{*} = \bigl((M-N)\bar{\alpha}^{(2)}, N\bigr)$, we set $\omega = 0 $ and $\bar{\delta}=\bar{\alpha}^{(2)}$, which - after recalling that we label the receivers so that $\bar{\alpha}^{(1)} \geq \bar{\alpha}^{(2)}$ - gives $ \Delta_{\text{com}} =  \bigl( N- (M-N)\bar{\alpha}^{(2)}  \bigr)$, which in turn gives (cf.~\eqref{eq:DoFd1},\eqref{eq:DoFd2})
\begin{align}
d_1 &= (M-N)\bar{\delta} + N( \bar{\delta} -  \bar{\alpha}^{(2)})^{+} =   (M  -  N)\bar{\alpha}^{(2)} \label{eq:Bd1Gen1}\\
d_2 &=  (M-N)\bar{\delta} + N( \bar{\delta} -  \bar{\alpha}^{(1)})^{+} + \Delta_{\text{com}} \label{eq:Bd2Gen1}\\
    &=  (M-N)\bar{\alpha}^{(2)} + N( \bar{\alpha}^{(2)} -  \bar{\alpha}^{(1)})^{+} + \bigl( N- (M-N)\bar{\alpha}^{(2)} - N( \bar{\alpha}^{(2)} - \bar{\alpha}^{(1)})^{+}  \bigr) = N. \label{eq:Bd2Gen2}
\end{align}

To achieve $D^{*} =\bigl(N, (M-N)\bar{\alpha}^{(1)}\bigr)$, we set $\omega = \frac{N( 1 +  \bar{\alpha}^{(1)} +\bar{\alpha}^{(2)})-(M+N) \bar{\delta} }{\Delta_{\text{com}} } = 1$ and $\bar{\delta}=\bar{\alpha}^{(1)}$, which gives $
 \Delta_{\text{com}} = \bigl( N- (M-N)\bar{\alpha}^{(1)} - N( \bar{\alpha}^{(1)} - \bar{\alpha}^{(2)})^{+}  \bigr) = N-M\bar{\alpha}^{(1)}+N\bar{\alpha}^{(2)}$,
which in turn gives (cf.~\eqref{eq:DoFd1},\eqref{eq:DoFd2})
\begin{align}
d_1 &= (M-N)\bar{\delta} + N( \bar{\delta} -  \bar{\alpha}^{(2)})^{+} + \omega\Delta_{\text{com}} =  (M -N)\bar{\alpha}^{(1)}  + N( \bar{\alpha}^{(1)} -  \bar{\alpha}^{(2)})^{+} +  \Delta_{\text{com}} = N\label{eq:Dd1Gen2}\\
d_2 &=  (M-N)\bar{\delta} + N( \bar{\delta} -  \bar{\alpha}^{(1)})^{+} + (1-\omega)\Delta_{\text{com}} =  (M -N)\bar{\alpha}^{(1)} \label{eq:Dd2Gen2}.
\end{align}
\vspace{3pt}
\emph{Case 2 - \eqref{eq:betaSuf} and \eqref{eq:alpha1Cond} and \eqref{eq:alpha12CondNot} (points $D^{*},B^{*},C^{*}$):}

Again under the condition of \eqref{eq:betaSuf}, we now consider the case where \eqref{eq:alpha1Cond} and \eqref{eq:alpha12CondNot} hold, and seek to achieve points $D^{*},B^{*},C^{*}$. For points $D^{*},B^{*}$, we can use the same parameters $\omega,\deltabar$ that we used before to achieve these same points (for
$B^{*}$ we set $\bar{\delta}=\bar{\alpha}^{(2)},\omega = 0$, and for $D^{*}$ we set $\bar{\delta}=\bar{\alpha}^{(1)}, \omega = \frac{N( 1 +  \bar{\alpha}^{(1)} +\bar{\alpha}^{(2)})-(M+N) \bar{\delta} }{\Delta_{\text{com}} } = 1$).
To achieve point $C^{*}$, we need to set $\omega = 0$ and $\deltabar = \frac{  N( 1 +  \bar{\alpha}^{(1)} +\bar{\alpha}^{(2)}) }{ M+N   }$ which, as before, gives
$ d_1 = (M-N)\bar{\delta} + N( \bar{\delta} -  \bar{\alpha}^{(2)})^{+} =  \frac{MN}{M+N}(1+\bar{\alpha}^{(1)}-\frac{N}{M}\bar{\alpha}^{(2)})$ and $d_2 =  (M-N)\bar{\delta} + N( \bar{\delta} -  \bar{\alpha}^{(1)})^{+} + \Delta_{\text{com}}   =  \frac{MN}{M+N}(1+\bar{\alpha}^{(2)}-\frac{N}{M}\bar{\alpha}^{(1)})$.\\[3pt]
\emph{Case 3 - \eqref{eq:betaSuf} and \eqref{eq:alpha1CondNot} (points $B^{*},A^{*}$):}

Again given \eqref{eq:betaSuf}, we now move to the case where \eqref{eq:alpha1CondNot} holds, and seek to achieve points $B^{*}$ and $A^{*}$.  To achieve $B^{*}$ we can use the same parameters as before, and thus set $\bar{\delta}=\bar{\alpha}^{(2)},\omega = 0$. To achieve $A^{*} = (N,\frac{(M-N)N( 1 + \bar{\alpha}^{(2)})}{M})$ we simply set $\deltabar = \frac{  N( 1 +  \bar{\alpha}^{(2)})}{ M   }$ and $\omega = \frac{  N(1+\bar{\alpha}^{(2)})-M\deltabar  }{ \Delta_{\text{com}} } = 1.$\\[-2pt]

We now focus on the DoF points in the inner bound of Proposition~\ref{prop:MIMOgenCSITInner}, corresponding to the setting where \eqref{eq:betaSufNot} holds, rather than \eqref{eq:betaSuf}.
In addition to the aforementioned points $D^{*}$ and $B^{*}$, we will seek to achieve the new points
\bea
E & = & \bigl(M\min \{\bar{\beta}^{(1)},\bar{\beta}^{(2)}\} -N\bar{\alpha}^{(2)}, \  N\bar{\alpha}^{(2)} +N(1-\min \{\bar{\beta}^{(1)},\bar{\beta}^{(2)}\}) \bigr) \\
F & = &  \bigl(N\bar{\alpha}^{(1)} +N(1-\min \{\bar{\beta}^{(1)},\bar{\beta}^{(2)}\}), \ M\min \{\bar{\beta}^{(1)},\bar{\beta}^{(2)}\} -N\bar{\alpha}^{(1)} \bigr) \\
G & = &  \bigl(N, \ (M-N)\min \{\bar{\beta}^{(1)},\bar{\beta}^{(2)}\}\bigr).
\eea
Before proceeding, we note that under~\eqref{eq:betaSufNot}, the bound on $\deltabar$ in~\eqref{eq:bardelta} now becomes
\begin{align} \label{eq:bardeltaLowQualDelay}
\bar{\delta}  \leq \min \{\bar{\beta}^{(1)},\bar{\beta}^{(2)}\}.
\end{align}
We proceed with the different cases, and now additionally consider the cases where
\begin{align}
\min \{ \bar{\beta}^{(1)}, \bar{\beta}^{(2)}\} & \geq  \bar{\alpha}^{(1)} \label{eq:betaAlpha1Suf}\\
\min \{ \bar{\beta}^{(1)}, \bar{\beta}^{(2)}\} & <   \bar{\alpha}^{(1)}.\label{eq:betaAlpha1SufNot}
\end{align}
\emph{Case 4a - \eqref{eq:betaSufNot} and \eqref{eq:alpha1Cond} and \eqref{eq:betaAlpha1Suf} (points $D^{*},B^{*},E,F$):}

To achieve $D^{*},B^{*}$ we use the same parameters as before, where for
$B^{*}$ we set $\omega = 0,\bar{\delta}=\bar{\alpha}^{(2)}$, and for $D^{*}$ we set $\bar{\delta}=\bar{\alpha}^{(1)}, \omega = 1 $, all of which satisfy the conditions in \eqref{eq:bardeltaLowQualDelay} and \eqref{eq:betaAlpha1Suf}.

To get point $E$, we set $\omega = 0$ and $\deltabar = \min \{\bar{\beta}^{(1)},\bar{\beta}^{(2)}\}$, and calculate that
\begin{align} d_1 &= (M-N)\bar{\delta} + N( \bar{\delta} -  \bar{\alpha}^{(2)})^{+} = M\min \{\bar{\beta}^{(1)},\bar{\beta}^{(2)}\} -N\bar{\alpha}^{(2)}\label{eq:Eid1Gen1}   \\
d_2 &=  (M-N)\bar{\delta} + N( \bar{\delta} -  \bar{\alpha}^{(1)})^{+} + \Delta_{\text{com}} \label{eq:Eid2Gen1}\\
&=  (M-N)\bar{\delta} + N( \bar{\delta} -  \bar{\alpha}^{(1)})^{+}  +   \bigl( N- (M-N)\bar{\delta} - N( \bar{\delta} - \bar{\alpha}^{(1)})^{+} - N( \bar{\delta} - \bar{\alpha}^{(2)})^{+}  \bigr)  \\
&=      N - N( \bar{\delta} - \bar{\alpha}^{(2)})^{+}  =    N\bar{\alpha}^{(2)} +N(1-\min \{\bar{\beta}^{(1)},\bar{\beta}^{(2)}\}) .  \label{eq:Eid2Gen2}
\end{align}
To get point $F$, we set $\deltabar = \min \{\bar{\beta}^{(1)},\bar{\beta}^{(2)}\}$ and $\omega = 1$, and calculate that
\begin{align} d_1 &= (M-N)\bar{\delta} + N( \bar{\delta} -  \bar{\alpha}^{(2)})^{+} + \omega\Delta_{\text{com}} \\
  & =  (M -N) \min \{\bar{\beta}^{(1)},\bar{\beta}^{(2)}\} + N( \min \{\bar{\beta}^{(1)},\bar{\beta}^{(2)}\} -  \bar{\alpha}^{(2)})^{+} + \Delta_{\text{com}}\\
  & =   N\bar{\alpha}^{(1)} +N(1-\min \{\bar{\beta}^{(1)},\bar{\beta}^{(2)}\})    \\
d_2 &=  (M-N)\bar{\delta} + N( \bar{\delta} -  \bar{\alpha}^{(1)})^{+} + (1-\omega)\Delta_{\text{com}}  =  M\min \{\bar{\beta}^{(1)},\bar{\beta}^{(2)}\} -N\bar{\alpha}^{(1)}.
\end{align}
\vspace{3pt}
\emph{Case 4b - \eqref{eq:betaSufNot} and \eqref{eq:alpha1Cond} and \eqref{eq:betaAlpha1SufNot} (points $B^{*},E,G$):}

Again under \eqref{eq:betaSufNot}, we now consider the case where \eqref{eq:alpha1Cond} and \eqref{eq:betaAlpha1SufNot} hold, and seek to achieve points $B^{*},E$ and $G$.
To achieve $B^{*}$ we set as before $\omega = 0,\bar{\delta}=\bar{\alpha}^{(2)}$, and for $E$ we set as before $\omega = 0$ and $\deltabar = \min \{\bar{\beta}^{(1)},\bar{\beta}^{(2)}\}$, both in accordance with the conditions in \eqref{eq:bardeltaLowQualDelay} and \eqref{eq:betaAlpha1SufNot}.

To get point $G=  \bigl(N, \ (M-N)\min \{\bar{\beta}^{(1)},\bar{\beta}^{(2)}\}\bigr) $, we simply set $\deltabar = \min \{\bar{\beta}^{(1)},\bar{\beta}^{(2)}\}$ and $\omega = 1$, and the calculations follow immediately.\\[3pt]
\emph{Case 4c - \eqref{eq:betaSufNot} and \eqref{eq:alpha1CondNot} (points $B^{*},E,G$):}

In the last case where \eqref{eq:betaSufNot} and \eqref{eq:alpha1CondNot} hold, we can achieve points $B^{*},E,G$ using the same parameters as above.

Finally the DoF regions in the theorem and proposition are achieved by time sharing between the proper DoF corner points.

\subsection{Modifications for the IC \label{sec:scheme_IC}}
We here briefly describe the modifications that adapt our scheme to the IC setting.  In terms of notation, the role of $\hHta \defeq \hHa[t]$
is taken by $\hHtab \defeq \hHab[t]$, of $\hHtb \defeq \hHb[t]$ by $\hHtba \defeq \hHba[t]$, of $\cHta \defeq \hHa[t+\eta]$ by $\cHtab \defeq \hHab[t+\eta]$, and the role of $\cHtb \defeq \hHb[t+\eta] $ is taken by $\cHtba \defeq \hHba[t+\eta]$.

Many of the steps follow from the BC, with the main difference being that now the common symbols must be transmitted by two independent transmitters. For that we change the structure of the signaling (cf.~\eqref{eq:TxX1Ph1}) and now consider that at time $t\in \Bc_s$ (phase~$s$), transmitter~1 sends
\begin{align} \label{eq:TxX1Ph1IC1}
\xv_{t}^{(1)} =  \Wm_t^{(1)}  \cv_t^{(1)} + \Um_t \av_t + \Um_t^{'} \av^{'}_t
\end{align}
and transmitter~2 sends
\begin{align} \label{eq:TxX1Ph1IC2}
\xv_{t}^{(2)} =  \Wm_t^{(2)}  \cv_t^{(2)} + \Vm_t \bv_t + \Vm_t^{'} \bv^{'}_t
\end{align}
where $\cv_t^{(i)}\in \CC^{ M \times 1}$ is the common information vector sent by transmitter~$i$ ($i=1,2$), where
$\Um_t$ is orthogonal to $\hHtba$, $\Vm_t$ is orthogonal to $\hHtab$, and where $\Wm_t^{(1)},\Um_t^{'},\Wm_t^{(2)},\Vm_t^{'}$ are randomly picked precoding matrices. Finally $\av_t, \av^{'}_t, \bv_t,\bv^{'}_t$ accept the same rate and power allocation previously described in~\eqref{eq:RPowerX1Ph1}.

In the above, the common symbol vectors $\{\cv_{\Bc_{s,t} }^{(1)}\}_{t=1}^T$ convey information on the (quantized version of the) interference $\iota^{(2)}_t  \defeq \Hba (\Um_t \av_t + \Um_t^{'} \av_t^{'})$, (cf.~\eqref{eq:c2barsg}) accumulated during phase~$(s-1)$.  These symbols will carry
\begin{align}
\bigl (\omega T \Delta_{\text{com}} + TN(\bar{\delta}-\bar{\alpha}^{(2)})^{+} \bigr)\log P -T o(\log P) \label{eq:ratejointcIC1}
\end{align}
bits, where $\Delta_{\text{com}}$ is defined in~\eqref{eq:com_inter}, and where $\omega \in [0,1]$ will be set depending on the target DoF point.
Similarly $\{\cv_{\Bc_{s,t} }^{(2)}\}_{t=1}^T$ will carry the
\begin{align}
    \bigl ( (1-\omega)T\Delta_{\text{com}} + TN(\bar{\delta}-\bar{\alpha}^{(1)})^{+} \bigr)\log P - T o(\log P) \label{eq:ratejointcIC2}
\end{align}
bits of information corresponding to $\{\iota^{(1)}_t \}_{t\in \Bc_{s-1}}$ where we recall that $\iota^{(1)}_t  = \Hab(\Vm_t \bv_t + \Vm_t^{'} \bv_t^{'})$ (cf.~\eqref{eq:c2barsg}). Jointly $\{\cv_{\Bc_{s,t} }^{(1)},\cv_{\Bc_{s,t} }^{(2)}\}_{t=1}^T$ will carry $T(N-(M-N)\bar{\delta}) \log P -o(\log P)$ bits, which matches the amount in the BC setting (cf. \eqref{eq:ratejointc}).

Decoding is similar to the case of the BC, except that now the corresponding MIMO MAC (for receiver~1) takes the form
\begin{align} \label{eq:firstMIMOx1IC}
\Bmatrix{ \ya[{\Bc_{s,1}}]-\bar{\check{\iota}}^{(1)}_{\Bc_{s,1}}
          \\ \bar{\check{\iota}}^{(2)}_{\Bc_{s,1}}
                    \\   \vdots
                    \\  \ya[{\Bc_{s,T}}]-\bar{\check{\iota}}^{(1)}_{\Bc_{s,T}}
          \\  \bar{\check{\iota}}^{(2)}_{\Bc_{s,T}} }   &=
\Bmatrix{  \begin{array}{ccc}
\Haa[{\Bc_{s,1}}] \Wm_{\Bc_{s,1}}^{(1)} &        &   \\
 \mathbf{0}        &        &   \\
                   &  \ddots  & \\
                   &        & \Haa[{\Bc_{s,T}}]\Wm_{\Bc_{s,T}}^{(1)} \\
                                     &        &  \mathbf{0}   \end{array} \!}  \! \Bmatrix{\! \cv_{\Bc_{s,1}}^{(1)} \\ \vdots \\  \cv_{\Bc_{s,T}}^{(1)}\!}
+\Bmatrix{  \begin{array}{ccc}
\Hab[{\Bc_{s,1}}]\Wm_{\Bc_{s,1}}^{(2)} &        &   \\
 \mathbf{0}        &        &   \\
                   &  \ddots & \\
                   &        & \Hab[{\Bc_{s,T}}]\Wm_{\Bc_{s,T}}^{(2)} \\
                                     &        &  \mathbf{0}   \end{array} }  \Bmatrix{ \cv_{\Bc_{s,1}}^{(2)} \\ \vdots \\  \cv_{\Bc_{s,T}}^{(2)}} \nonumber\\
&\quad +\Bmatrix{  \begin{array}{ccc}
\Bmatrix{  \Haa[{\Bc_{s,1}}] \\   \cHtba[{\Bc_{s,1}}] }  \Bmatrix{\!  \Um_{\Bc_{s,1}}  \    \Um_{\Bc_{s,1}}^{'} } &        &   \\
                   & \ddots  & \\
                                     &        &  \Bmatrix{  \Haa[{\Bc_{s,T}}] \\   \cHtba[{\Bc_{s,T}}] } \! \!\Bmatrix{ \Um_{\Bc_{s,T}}  \    \Um_{\Bc_{s,T}}^{'} }  \end{array} } \! \!\Bmatrix{ \av_{\Bc_{s,1}} \\  \av_{\Bc_{s,1}}^{'} \\  \vdots \\  \av_{\Bc_{s,T}} \\  \av_{\Bc_{s,T}}^{'} }
 + \Bmatrix{\! \tilde{\zv}^{'(1)}_{\Bc_{s,1}} \\   -\tilde{\iota}^{(2)}_{\Bc_{s,1}}  \\  \vdots \\  \tilde{\zv}^{'(1)}_{\Bc_{s,T}} \\  -\tilde{\iota}^{(2)}_{\Bc_{s,T}}  \!}
\end{align}
where the effective noise term at the end can be shown to have bounded power. As before, the receivers recover the signals at the rates described in \eqref{eq:RPowerX1Ph1}, \eqref{eq:ratejointcIC1}, \eqref{eq:ratejointcIC2} (cf.~\cite{CT:06}).

\section{Conclusions} \label{sec:conclu}
The work, extending on recent work on the MISO BC, considered the symmetric MIMO BC and MIMO IC, and made progress towards establishing and meeting the tradeoff between performance, and feedback timeliness and quality.  Considering a general CSIT process, the work provided simple DoF expressions that reveal the role of the number of antennas in establishing the feedback quality associated to a certain DoF performance.



\end{document}